# NOUVELLES PROBLEMATIQUES POSEES PAR LE CYCLAGE THERMO-MECANIQUE EN CAVITES SALINES

*QUESTIONS RAISED BY THERMO-MECHANICAL CYCLING IN SALT CAVERNS*

Cyrille PELLIZZARO[1], Pierre BEREST[2], Benoit BROUARD[3], Mehdi KARIMI-JAFARI[4]

*1 Storengy, Bois-Colombes, France*
*2 Ecole Polytechnique, Palaiseau, France*
*3 Brouard Consulting, Paris, France*
*4 Géostock, Rueil-Malmaison, France*

**RÉSUMÉ** — Le stockage de gaz en cavités salines a principalement été développé pour le stockage saisonnier (injection de gaz en été lorsque la demande est faible, soutirage en hiver lorsque la demande est forte). Ainsi les cavités subissent en général un cycle par an et les débits de gaz associés sont modérés. Lors des études géomécaniques conduites pour s'assurer de la stabilité des cavités, on étudie l'impact de la pression du gaz en cavité sur le massif de sel environnant, avec la particularité que le sel est un matériau fluant (élasto-visco-plastique). Aujourd'hui les besoins en énergie font évoluer le stockage en cavité vers des modes d'exploitation plus agressifs, avec davantage de cycles par an et des débits beaucoup plus élevés. En plus de subir des variations de pression, le gaz subit des variations de température qui ne sont plus négligeables et qui peuvent avoir un impact sur l'intégrité de la paroi de la cavité. Ce nouveau mode d'exploitation soulève ainsi de nouvelles questions concernant les études de stabilité à mener, des lois de comportement du sel aux procédures d'essais en laboratoire ainsi qu'aux critères à utiliser pour s'assurer de la stabilité à court et à long terme des cavités salines.

**ABSTRACT** — Storage of natural gas in salt caverns had been developed mainly for seasonal storage, resulting in a small number of yearly pressure cycles and moderate gas-production rates. The needs of energy traders are changing towards more aggressive operational modes. The "high-frequency cycling" operation of salt caverns raises questions concerning the effects of frequently repeated and intense mechanical and thermal loading. These questions concern the constitutive creep laws for salt, laboratory test procedures, criteria to be used at the design stage to provide operability, and the long-term integrity of the underground salt caverns.

## 1. Introduction

Le sel gemme est une roche particulièrement adaptée au stockage d'hydrocarbures : il est soluble dans l'eau (cette propriété est mise à profit pour le creusement des cavités par dissolution) et sa perméabilité est très faible (typiquement $10^{-18}$ -$10^{-21}$ m$^2$).

Le volume d'une cavité saline est généralement compris entre 10 000 m$^3$ et 1 000 000 m$^3$ et sa profondeur entre 500 m et 2000 m. Elle est reliée à la surface par un puits de type pétrolier (diamètre de l'ordre de 20 cm), permettant au gaz d'être injecté et soutiré à souhait par compression / détente.

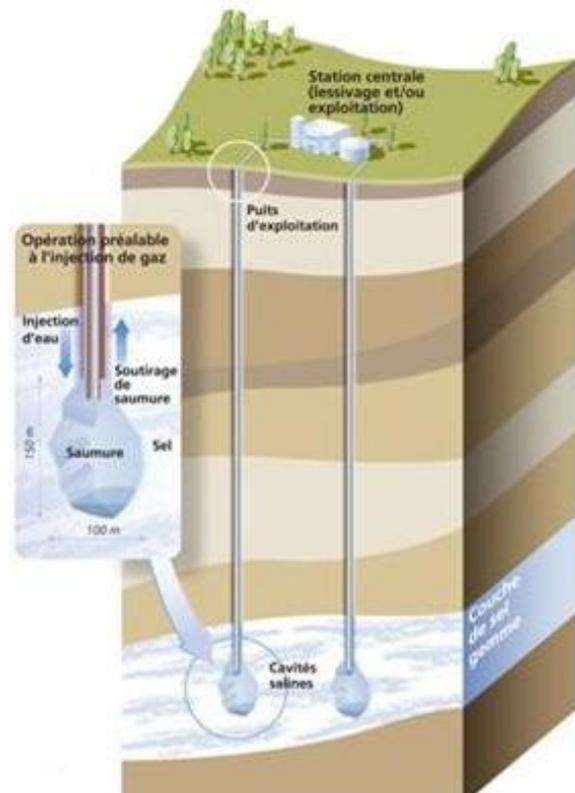

Initialement développé pour répondre à une consommation de gaz saisonnière, le stockage en cavités salines doit aujourd'hui s'adapter à un fort besoin de flexibilité. En plein essor, les centrales électriques à gaz nécessitent d'être approvisionnées au cours de cycles intra-hebdomadaires (jours ouvrés / week-end) voire intra-journaliers (heures pleines / heures creuses). Le stockage d'air comprimé en cavité saline est également envisagé pour stocker l'excédent électrique produit pendant les heures creuses (par des énergies intermittentes telles que l'énergie éolienne) et le restituer par détente pendant les heures pleines.

Les études géomécaniques conduites pour s'assurer de la stabilité à court et long terme des cavités doivent donc intégrer deux nouvelles problématiques, celle de l'augmentation de la fréquence des cycles d'injection / soutirage et celle d'un cyclage non plus uniquement mécanique mais aussi thermique, les variations de température ne pouvant plus être considérées comme négligeables au cours de ces nouveaux modes d'exploitation. Ainsi de nouvelles questions sont soulevées concernant les études de stabilité à mener, des lois de comportement des matériaux et critères de stabilité à utiliser aux procédures d'essais conduits en laboratoire.

## 2. Nouveaux scénarios d'exploitation : conséquences en cavité

### 2.1. Variations de pression du gaz

Une cavité saline exploitée de manière saisonnière subit des décompressions moyennes de 0,2-0,3 MPa/jour (jusqu'à 1-1,5 MPa/jour ponctuellement) tandis que les nouveaux scénarios d'exploitation impliquent des soutirages à 1-1,5 MPa/jour minimum, voire beaucoup plus élevés (0,5-1,5 MPa/heure) mais pour une plage de pression moindre (chaque cavité saline est exploitée entre une P$_{max}$ et une P$_{min}$ fixées, assurant sa stabilité pour une exploitation saisonnière) (Figure 1).

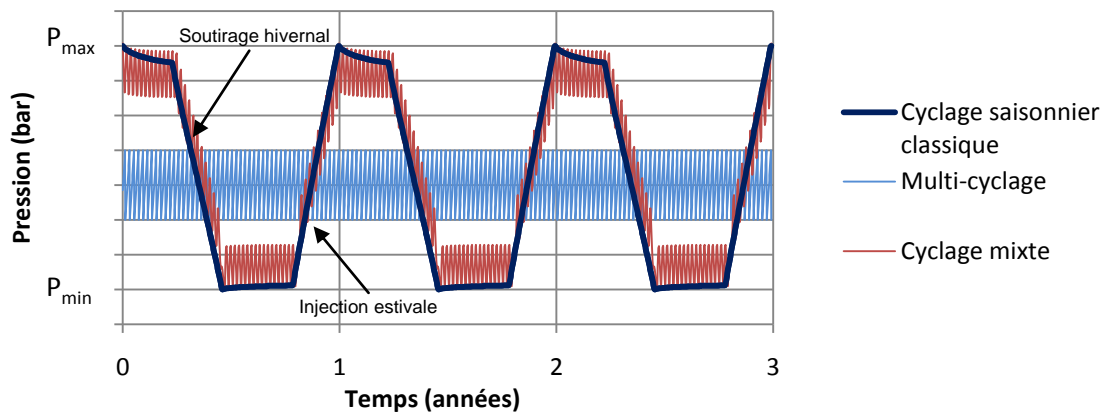

Figure 1. Cycle de pression en cavité suivant le mode d'exploitation retenu

## 2.2. *Variations de température du gaz*

La température du gaz en cavité est beaucoup moins prévisible que sa pression car elle dépend de la température du gaz injecté, de la compression / décompression qu'on lui fait subir et également des échanges thermiques avec le massif. Une cavité saline profonde (1500 m) exploitée de manière saisonnière subit des variations de température moyennes de 0,25-0,5°C/jour (jusqu'à 5°C/jour ponctuellement), menant à des amplitudes de 30-40°C lors des cyclages annuels. Augmenter la fréquence des cycles d'exploitation a deux conséquences :

- La première est d'augmenter les taux de variation de température en cavité, induisant un gradient thermique en paroi plus élevé, soit une épaisseur de sel affectée par les changements de température plus restreinte. En effet lors d'un cycle hebdomadaire la température est modifiée sur environ 2 m de sel tandis que lors d'un cycle annuel la température est modifiée sur une dizaine de mètres (Figure 2).

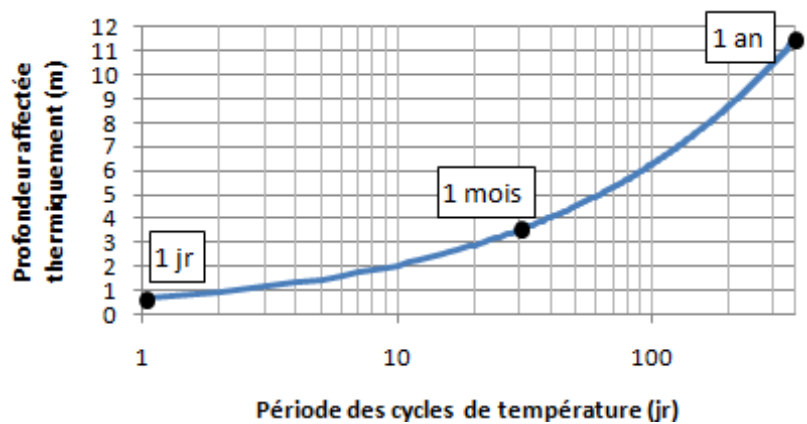

Figure 2. Profondeur de pénétration ($dT > dT_{paroi}/10$) de variations thermiques sinusoïdales à la paroi en fonction de leur période pour une cavité de rayon 50 m et une diffusivité thermique du sel de $3.10^{-6}$ m$^2$.s$^{-1}$ (d'après Lestringant *et al.*, 2010)

- La seconde est d'augmenter l'amplitude des cycles de température en cavité car les échanges thermiques avec le massif ont moins de temps pour s'effectuer. Ainsi un cyclage mensuel peut impliquer des variations de température en cavité 3 fois plus grandes qu'un cyclage annuel (Figure 3).

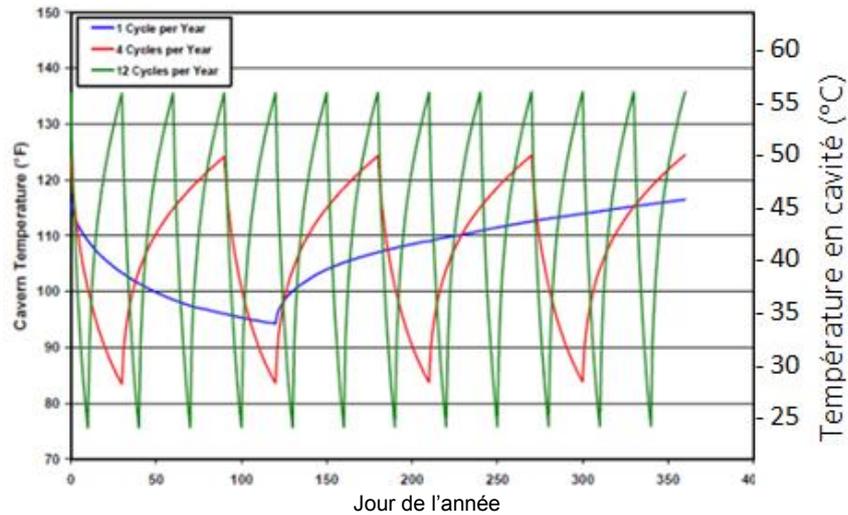

Figure 3. Variations de température en cavité suivant le nombre de cycles par an (Niedland, 2008)

## 3. Conséquence sur la stabilité de la cavité

### 3.1. Modélisation géomécanique

Le calcul des contraintes dans le massif s'effectue en général numériquement. Prenons l'exemple d'une cavité peu profonde subissant une décompression de type « cyclage mixte » (Figure 4) :

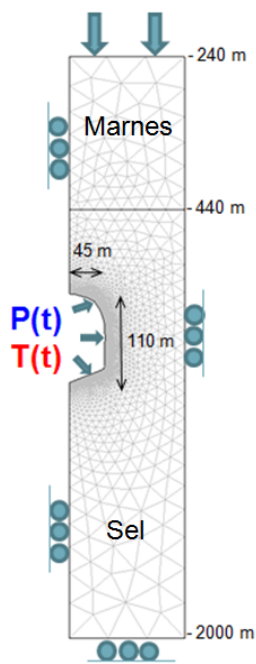

- Volume (creusé) de la cavité : 600 000 m$^3$
- Profondeur moyenne : 600 m
- Etat initial des terrains à mi-cavité :

$$\begin{cases} \sigma_r = \sigma_\varphi = \sigma_z = -14,7 \text{ MPa} \quad (\text{convention} < 0 \text{ en compression}) \\ T = 26°C \end{cases}$$

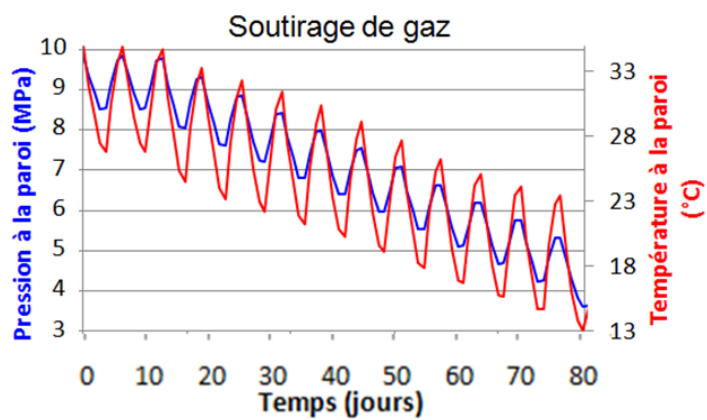

Figure 4. Modélisation géomécanique d'une cavité saline (soutirage de gaz)

Le sel est un matériau fluant susceptible de tolérer de grandes déformations sans se rompre. On le modélise ici par un matériau élasto-viscoplastique de Lemaître (1970) (à 5 paramètres $\alpha, \beta, K_{Tref}, T_{ref}, B$). La déformation $\underline{\underline{\varepsilon}}$ est liée à la contrainte $\underline{\underline{\sigma}}$ et son déviateur $\underline{\underline{s}}$ par la relation :

$$\underline{\underline{\dot{\varepsilon}}} = \frac{1+\nu}{E}\underline{\underline{\dot{\sigma}}} - \frac{\nu}{E}tr\left(\underline{\underline{\dot{\sigma}}}\right)\underline{\underline{1}} + \alpha_l \dot{T} \underline{\underline{1}} + \frac{3}{2}\frac{\underline{\underline{s}}}{\sqrt{3J_2}}\frac{d\zeta^\alpha}{dt}.10^{-6} e^{\left(-B\left(\frac{1}{T}-\frac{1}{T_{ref}}\right)\right)} \quad avec \quad \dot{\zeta} = \left(\frac{\sqrt{3J_2}}{K_{Tref}}\right)^{\frac{\beta}{\alpha}} \quad (1)$$

La température joue un rôle via la dilatation thermique ($\alpha_l = 3.10^{-5}$ K$^{-1}$) ainsi que via le facteur exponentiel d'Arrhenius qui accentue la déformation viscoplastique quand la température du milieu augmente (de 1 à 2% par °C).

### 3.2. Critères de stabilité

Compte tenu de la ductilité du sel, on se focalise davantage sur des critères en contraintes qu'en déformation, les principaux étant les suivants :

#### 3.2.1. Absence de traction

La résistance à la traction du sel est très faible (de l'ordre de 2 MPa). Le critère usuel est donc de s'assurer que le chargement appliqué à la paroi de la caverne n'induit aucune contrainte de traction dans le massif environnant.

#### 3.2.2. Absence de dilatance

Lorsqu'un échantillon de sel est soumis à une contrainte de cisaillement élevée (par rapport au confinement), on observe une augmentation de volume interprétée comme l'apparition de micro-fractures au sein de celui-ci. La dilatance peut ainsi être considérée comme un signe d'endommagement.

Les critères de dilatance sont généralement formulés en terme d'invariants des contraintes faisant intervenir la contrainte déviatorique ($J_2$) et la contrainte moyenne ($I_1/3$), par exemple (Spiers *et al.*, 1988) :

$$\sqrt{J_2} < -0{,}27 I_1 + 1{,}9 \text{ MPa} \quad (2)$$

$$avec \quad J_2 = \frac{1}{6}\left((\sigma_1-\sigma_3)^2 + (\sigma_2-\sigma_1)^2 + (\sigma_3-\sigma_2)^2\right) \quad et \quad I_1 = \sigma_1+\sigma_2+\sigma_3$$

### 3.3. Résultats

La Figure 5 présente les contraintes calculées au voisinage de la cavité suite au soutirage (ayant amené sa pression de 10 MPa à 3,5 MPa et sa température de 35°C à 14°C), avec et sans prise en compte de la température. La prise en compte de la température modifie les contraintes tangentielles dans les premiers mètres à la paroi. Suite au refroidissement, celles-ci entrent en traction sur quelques dizaines de centimètres.

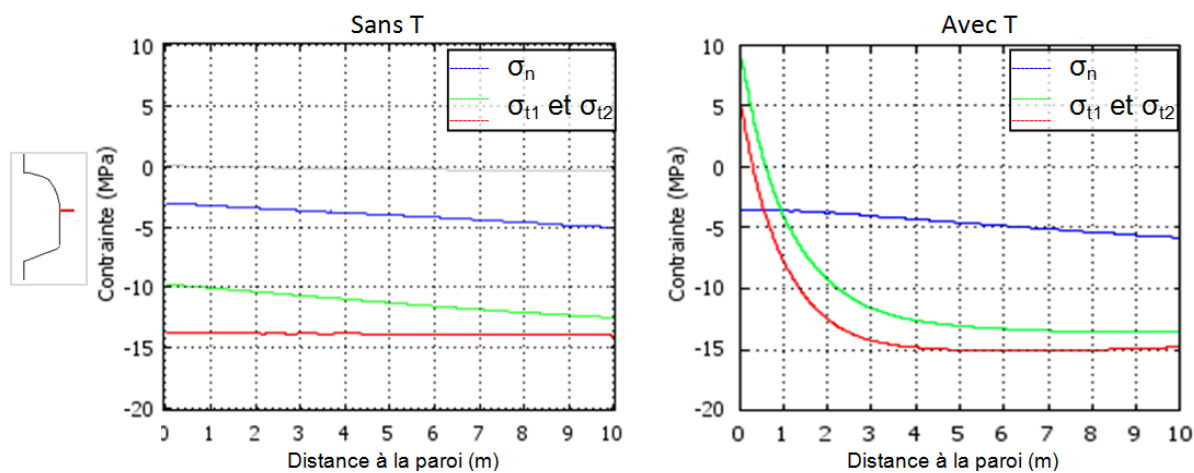

Figure 5. Contraintes normales et tangentielles en fonction de la distance à la paroi en fin de soutirage (calcul méca vs. calcul thermo-méca)

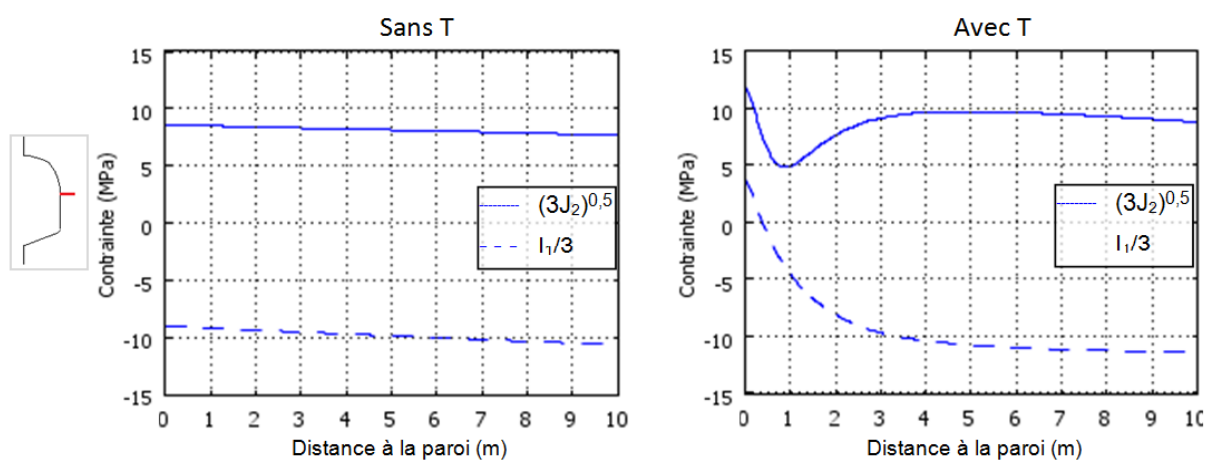

Figure 6. Contrainte déviatorique de Mises $(3J_2)^{0,5}$ et contrainte moyenne ($I_1/3$) en fonction de la distance à la paroi en fin de soutirage
(calcul méca vs. calcul thermo-méca)

La contrainte déviatorique est quant à elle relativement peu affectée par la température mais la contrainte moyenne l'est fortement (Figure 6) (déconfinement en raison des tractions), faisant apparaître de la dilatance à la paroi avec le critère (2).
Ainsi, plus les variations de température en cavité sont importantes, plus le risque de générer (lors des refroidissements) de la traction tangentielle et de la dilatance, donc de violer les critères usuels de stabilité, dans une zone proche de la paroi (de l'ordre métrique) est important.
Sur cet exemple le multicyclage n'a pas d'impact mécanique direct. Son principal effet est d'amplifier la variation de température globale en cavité donc d'augmenter les contraintes tangentielles obtenues en fin de soutirage.

*3.4. Discussion*

*3.4.1. Origine des contraintes thermiques*

L'ordre de grandeur des contraintes induites à la paroi par une variation de température dans le cas d'un milieu élastique est :

$$\Delta\sigma_n = 0 \quad \text{et} \quad \Delta\sigma_t = -\frac{E\alpha_l}{1-\nu}\Delta T \tag{3}$$

Les contraintes tangentielles varient donc d'environ 1 MPa/°C (compression lors d'un échauffement, traction lors d'un refroidissement). Ces surcontraintes induites par la température, ainsi que celles induites par la variation de pression, viennent s'ajouter à celles en place avant le soutirage. Compte tenu du comportement différé du sel, ces dernières dépendent de l'ensemble de l'historique d'exploitation (P&T), d'où la difficulté d'édicter des règles d'exploitation générales pour prévenir les tractions en paroi : une décompression admissible à un certain moment ne l'est pas forcément ultérieurement.

*3.4.2. Validité des critères et des lois utilisées*

De nombreuses lois de comportement existent pour le sel. Il est difficile de déterminer quelle est la plus pertinente lorsqu'un cyclage rapide en pression et en température est considéré. La loi utilisée doit au minimum comporter une phase de fluage transitoire ainsi qu'une dépendance à la température, deux propriétés que possède la loi de Lemaître.
Mellegard et Dusterloh (2012) montrent qu'un échantillon soumis à un chargement cyclique se déforme d'une manière semblable à un chargement constant pourvu qu'il ne soit pas dans la phase dilatante. Certains (Zander, 2010) suggèrent d'introduire dans la loi de comportement des variables internes d'endommagement.
Concernant le critère de dilatance, les résultats sont actuellement contradictoires. Bauer *et al.* (2010) observent de la dilatance après un certain nombre de cycles appliqués sur un échantillon alors que la limite de dilatance « statique » n'est pas atteinte. Arnold *et al.* (2011) quant à eux ne mettent pas en évidence de fatigue du matériau suite à un nombre important de cycles de pression (5000). Des investigations supplémentaires sont donc nécessaires, d'autant que la plupart des essais actuels sont conduits à température constante et ne font pas intervenir la vitesse de déformation, qui pourrait jouer sur la résistance du sel (Wallner, 1984).

*3.4.3. Conséquences sur la stabilité de l'ouvrage*

Si les tractions induites par le refroidissement sont excessives, des fractures sont générées à la paroi. Les tractions étant tangentielles, les éventuelles fractures sont perpendiculaires à la paroi donc moins enclines à générer de l'écaillage que des fractures tangentielles (Pellizzaro *et al.*, 2011). Il est également admis que les fissures ne se propagent pas indéfiniment, leur profondeur correspondant à peu près à celle de la zone en traction, elle-même correspondant à celle affectée par le refroidissement. Le retour d'expérience indique enfin que des cavités qui devraient subir des tractions à la paroi avec apparition de fissures, restent stables et ne subissent pas d'évolution de forme significative. Ces éléments vont donc plutôt dans le sens de la sécurité mais nécessitent d'être validés expérimentalement.

## 4. Conclusion

Lors d'un cyclage saisonnier, les variations de pression en cavité sont lentes et les transferts de chaleur entre le gaz et le massif de sel suffisants pour limiter les variations de température en cavité dues aux compressions / détentes. Lorsque la fréquence des cycles d'exploitation augmente, ces transferts thermiques ont moins de temps pour s'effectuer, induisant des variations de température en cavité plus importantes. Celles-ci modifient l'état de contraintes du sel et aboutissent en cas de refroidissement excessif à de la traction et de la dilatance dans le premier mètre à la paroi, violant ainsi deux critères classiquement utilisés lors des études de stabilité des cavités. De multiples éléments indiquent que cet endommagement reste localisé à la paroi donc que la stabilité globale de l'ouvrage n'est pas remise en cause. Ceci nécessite cependant d'être vérifié expérimentalement. Une réflexion doit enfin être menée sur les critères et lois à utiliser lors des études de conception car de tels cyclages rapides en pression et température sont amenés à devenir courants.